\documentclass[aps,prb,preprint,floatfix,raggedbottom]{revtex4-1} 

\usepackage{bm}
\usepackage{amsmath}
\usepackage{amsfonts}
\usepackage{amssymb}
\usepackage{graphicx}
\usepackage{setspace}
\usepackage{booktabs} 
\usepackage{ragged2e} 

\begin{document}

\title{Unraveling the origins of conduction band valley degeneracies in Mg$_2$Si$_{1-x}$Sn$_{x}$ thermoelectrics} 

\author{Chang-Eun \surname{Kim}}
\affiliation{Global E$^3$ Institute and Department of Materials Science and Engineering, Yonsei University, Seoul 120-749, Korea}

\author{Aloysius \surname{Soon}}
\email[Corresponding author. E-mail: ]{aloysius.soon@yonsei.ac.kr}
\affiliation{Global E$^3$ Institute and Department of Materials Science and Engineering, Yonsei University, Seoul 120-749, Korea}

\author{Catherine \surname{Stampfl}}
\affiliation{School of Physics, The University of Sydney, NSW 2006, Australia}
\date{\today}

\begin{abstract}
To better understand and enhance the thermoelectric efficiency of a new class of Mg-based thermoelectrics, using hybrid density-functional theory, we study the microscopic origins of valley degeneracies in the conduction band of the solid solution Mg$_2$Si$_{1-x}$Sn$_{x}$ and its constituent components -- namely, Mg$_2$Si and Mg$_2$Sn. In the solid solution of Mg$_2$Si$_{1-x}$Sn$_{x}$, the Mg sublattice and Si/Sn sublattice are expected to undergo either tensile or compressive strain. Interestingly, both tensile strain of Mg$_2$Si and compressive strain of Mg$_2$Sn enhance the conduction band valley degeneracy or band convergence, which has been strongly speculated as the electronic origin of the enhanced Seebeck coefficient in the Mg$_2$Si$_{1-x}$Sn$_{x}$ system. We also consider finite-temperature electronic band structures of these systems to account for high temperature effects. Our results clearly highlight and demonstrate the role of sublattice strain in the band valley degeneracy observed in Mg$_2$Si$_{1-x}$Sn$_{x}$.
\end{abstract}

\maketitle

\clearpage
\newpage

\begin{flushleft}
{\bf Graphical Abstract for Table-of-Content:} 
\end{flushleft}
\justify
The chemical and physical origin of the enhanced band valley degeneracy for a new class of Mg-based thermoelectrics Mg$_2$Si$_{1-x}$Sn$_{x}$ (MSS) is examined using temperature-broadened, orbital-projected band structure as calculated by hybrid density-functional theory (DFT-HSE06). For the MSS alloys, varying the composition of $x_{\rm Sn}$ modulates the chemical orbital nature of the conduction band edge states, and couples with the induced sublattice strain to further engineer the degree of conduction band valley degeneracies in these Mg-based alloys.
\vspace{1mm}
\begin{center}
\includegraphics[width=0.65\textwidth]{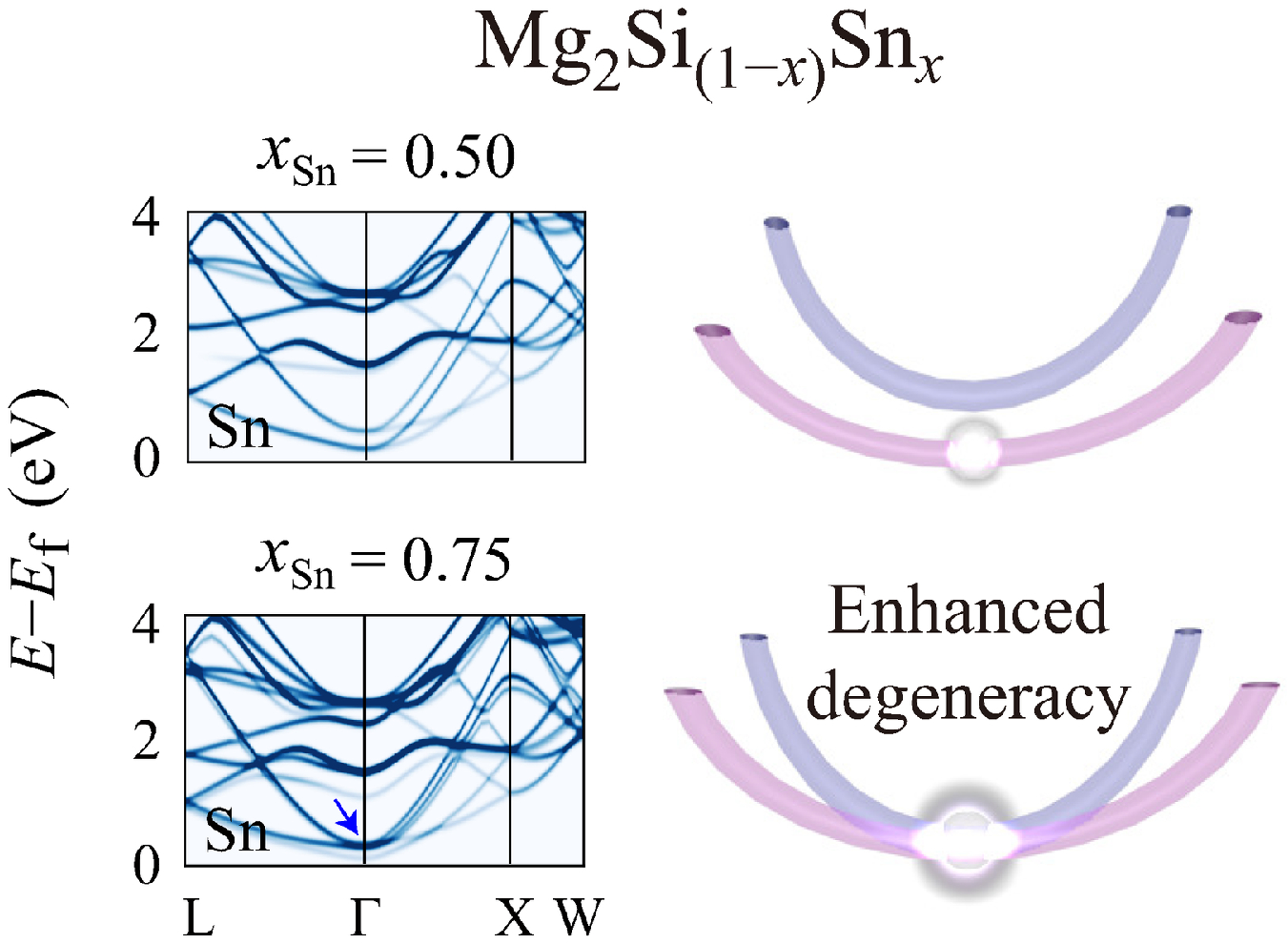}
\end{center}

\clearpage
\newpage

\section{Introduction}
Thermoelectric materials have a great potential to improve the clean energy sustainability of our modern society.\cite{heremans2008} These materials generate an electric voltage when a heat gradient is established inside, that can convert waste heat into useful electric potential. The efficiency of a thermoelectric materials can be described by its figure of merit. The figure of merit is given by $Z=\frac{\sigma S^2}{\kappa}$, where $S$, $\sigma$ and $\kappa$ are the Seebeck coefficient, electric conductivity, and the thermal conductivity, respectively. When $Z$ is multiplied by temperature, $T$, then $ZT$ becomes a dimensionless quantity that represents the intrinsic thermoelectric efficiency of a thermoelectric material at a specified temperature.

The $ZT$ value scales quadratically with the Seebeck coefficient, $S$. In the Mott equation,\cite{heremans2008} $S$ can be expressed as,
\begin{equation}
\label{eq-seebeck}
S=\frac{\pi^2}{3} \frac{k_{\rm B}}{q} k_{\rm B} T \left\{ \frac{1}{n} \frac{dn(E)}{dE} + \frac{1}{\mu} \frac{d\mu (E)}{dE} \right\}_{E=E_{\rm F}} \quad,
\end{equation}
where $k_{\rm B}$ is the Boltzmann constant, $q$ the carrier charge, $n(E)$ the carrier density at the energy $E$, $\mu(E)$ the mobility, and the $E_{\rm F}$ denotes the Fermi energy. As shown from Eq.\,\ref{eq-seebeck}, both $\frac{dn(E)}{dE}$ and $\frac{d\mu (E)}{dE}$ contribute positively to the Seebeck coefficient, leading to an overall improvement of the thermoelectric efficiency, $ZT$. 

Recently, an interesting observation reported for Mg$_2$Si$_{1-x}$Sn$_{x}$ (MSS) solid solution compounds was explained on this basis.\cite{liu2012} MSS alloys consist of Mg$_2$Si and Mg$_2$Sn, which are widely available, non-toxic, and cheap materials. MSSs share an optimal operational temperature range with PbTe thermoelectric materials (which is most efficient for power-generation at high temperatures), which makes it an attractive potential replacement material for PbTe composites when the toxicity of Pb, and the scarcity of Te, become an issue in conventional thermoelectric applications. 

\begin{figure}[tb!]
\center
\includegraphics[width=0.50\textwidth]{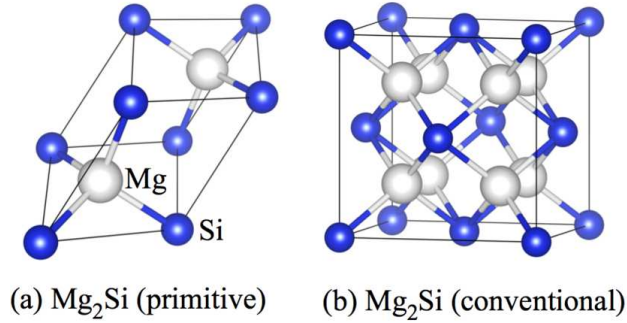} %0.38
\caption{(Color online) Crystal structure of the (a) primitive and (b) conventional unit cells of Mg$_2$Si. The Mg atoms in the primitive unit cell occupy the $8c$ sites of $Fm\bar{3}m$ space group which contributes to the band valley degeneracy at both the $\Gamma$ and $X$ points of the reciprocal unit cell.}
\label{figure-atomic}
\end{figure}

A significant enhancement in $ZT$ of MSSs by alloying an optimal proportion of Mg$_2$Si and Mg$_2$Sn has been reported.\cite{zaitsev2006,zhang2008a,gao2011,liu2012a} Initially, it was suggested that the relatively large effective mass of Mg$_2$Sn is responsible for increase in the $\frac{d\mu (E)}{dE}$ term of Eq.\,\ref{eq-seebeck}. Later, first-principles calculations in combination with experimental analyses suggested that the $\frac{dn(E)}{dE}$ term plays a prominant role in this alloy system,\cite{liu2012} and attributed the enhancement to the so-called ``convergence of conduction bands" by alloying. The observed enhancement was reproduced again with a different kind of dopant,\cite{liu2013} and a recent experimental study confirmed the Seebeck coefficient of MSS is improved by about 15--37\,\% as compared to previously reported Mg$_2$Si samples.\cite{dasgupta2014} 

Knowing the fundamental mechanism and principles behind this series of observations has significant implications because it could provides a rather straightforward and direct means to engineer the electronic structure of thermoelectric materials simply by alloying isovalent anions. Previously, an enhancement of the $\frac{dn(E)}{dE}$ term has been reported for the PbTe system, where a ``resonance level" in the observed density-of-states is established between the dopant and the semiconductor.\cite{heremans2008} 

The band structures of Mg$_2$Si and Mg$_2$Sn have inherently a high level of degeneracy due to their crystalline lattice structures (as shown in Fig.\,\ref{figure-atomic}). Pure Mg$_2$Si alone produces a modest thermoelectric efficiency of $ZT = 0.7$ at 775\,K.\cite{bux2011} On the other hand, Mg$_2$Sn does not show great thermoelectric power since it undergoes a semiconductor-to-metal transition at high temperatures. However, when Sn is doped to replace Si, the efficiency almost doubles at an optimal composition range of Sn ($x_{\rm Sn} = 0.5 \sim 0.7$). 

Considering the electronic band structure of MSS, it is suggested that the low-lying states of the conduction band are located close to each other, forming effectively degenerate levels in the solid solution at sufficiently high temperature.\cite{liu2012} A Sb-doped MSS with $x_{\rm Sn} = 0.6$ led to $ZT = 1.3$ at $T = 740$\,K, and a very similar performance was reproduced again with a Bi-doped MSS ($x_{\rm Sn} = 0.6$, with $ZT = 1.4$ at $T = 800$\,K).\cite{liu2013} Regardless of the choice of the dopants (Sb or Bi), the proportion of Sn acted as a dominant factor in the enhancement of $ZT$. 

To gain further insight into the experimental observations, a fundamental understanding of the electronic structure of these MSS composites is necessary, and there have been some recent studies in this direction. For instance, Tan {\it et al.}\cite{tan2012} calculated the band structure of MSS using density-functional theory (DFT) for various compositions of Sn using ultra-soft pseudopotentials and a semi-local exchange-correlation functional (GGA-PW91). They found that the two bands with different effective masses were found to be very close in energy, forming an effectively degenerate level for $x_{\rm Sn} = 0.625$ in the MSS solid solution. 

Using the semi-local generalized gradient approximation (GGA), the electronic structure of Mg$_2$Sn is calculated to be metallic.\cite{pulikkotil2012} This wrong electronic structure description and severe underestimation of the band gap as obtained by DFT-GGA is well known and expected. Given the known challenges in obtaining an accurate description of complex electronic band structures at the semi-local treatment of the DFT exchange-correlation ($xc$) functional, the study of these intricate band degeneracies may well require going beyond this semi-local approximation. Thus, in this work, we go beyond the GGA and employ a hybrid DFT $xc$ functional which is known to significantly improve the description of the electronic band structures of complex materials.\cite{heyd2004}

Here, we report the electronic band structures projected by the orbital character, detailing individual contributions from the atomic orbitals. In previous studies, the contribution of the filled 4$d$ state of Sn to the conduction edge of the band structure has not been discussed. In our projected band structure calculations, the presence of Sn 4$d$ states are clearly demonstrated, and this observation coincides with a similar enhancement found in the Co-Si-Ge alloy, where changes in the Co 3$d$ state was found to be closely correlated with an improved Seebeck coefficient of the $n$-type semiconductor.\cite{hsu2014} In addition, we address the possible origin of the ``convergence of bands" in MSS solid solutions as suggested in Ref.\,\onlinecite{liu2012}. In addition, the relative electronic energy levels in the band structure may well be affected by applied geometric strain as shown in PbTe-Se alloy.\cite{pei2011} Vegard's law suggests that the lattice parameter of the solid solution varies as a linear combination of the lattice parameters of the two constituent compounds. On this basis, the sublattice of Mg$_2$Si is expected to experience tensile strain, while that of Mg$_2$Sn will undergo compressive strain in the solid solution. Tensile and compressive strains in Mg$_2$Si and Mg$_2$Sn, respectively, are found to intensify the degree of conduction band valley degeneracy. 

As mentioned above, the enhanced band valley degeneracy in the conduction band has been suggested to be responsible for the improvement, however, details of the electronic band structure, and the effect of geometric structural variation such as lattice expansion/compression are presently not understood. In this work, we focus on the origin of the effectively degenerate conduction bands of Mg$_2$Si and Mg$_2$Sn, presenting detailed information of the projected band structure. 

\section{Methodology and computational approach}
We perform density-functional theory calculations as implemented in the Vienna {\it ab-initio} simulation package (VASP).\cite{kresse1996,kresse1996a} The kinetic energy cutoff for the planewave basis set is 400\,eV and the electron-ion interactions are represented using the projector augmented wave (PAW) potentials.\cite{blochl1994,kresse1999} A $\Gamma$-centered {\bf k}-point grid of 9$\times$9$\times$9 is used for geometric optimization. The total energy convergence criterion is 10$^{-5}$\,eV for electronic minimization steps, and 10$^{-4}$\,eV for ionic displacement steps. 

Both Mg$_2$Si and Mg$_2$Sn crystalize in the $Fm\bar{3}m$ fluorite cyrstal structure (as depicted in Fig.\,\ref{figure-atomic}) The crystalline structures of Mg$_2$Si and Mg$_2$Sn are calculated using various DFT $xc$ functionals and compared to experimental results. In particular, the HSE06,\cite{heyd2003,heyd2004,heyd2006,paier2006} PW91,\cite{perdew1992,perdew1993} PBE,\cite{perdew1996,perdew1997} PBEsol,\cite{perdew2008}, PBE+D2,\cite{grimme2006} and PBE+D3,\cite{grimme2010} functionals have been tested. The calculated lattice parameter of the primitive unit cells of Mg$_2$Si and Mg$_2$Sn are compared to experimental values, as detailed in Tab.\,\ref{table-lattice}. 

We calculated the band structures of pure Mg$_2$Si and Mg$_2$Sn for varying degrees of lattice strain, in order to understand the effect of the geometric lattice strain independently from other factors. Due to alloying, the Mg$_2$Si sublattice will experience tensile strain, while the Mg$_2$Sn sublattice will experience compressive strain.

To calculate electronic band structures of MSS solid solutions, we consider three representative alloy models. The MSS alloys are known to form stable homogeneous alloys in the range of $0.55 < x_{\rm Sn} < 1.0$ (where $x_{\rm Si} + x_{\rm Sn} = 1.0$),\cite{vives2014} and the optimal range of $x_{\rm Sn}$ is from 0.5 to 0.7 as reported from experimental measurements.\cite{liu2012a, dasgupta2014} To simulate these compositions, we use homogeneous solid solution models for $x = 0.5$ and 0.75, while an $x = 0.25$ model is used as a comparative case. A supercell which is eight times larger than the Mg$_2$Si primitive unit cell is created. There are 16 Mg atoms and 8 Si atoms in this $p(2\times2\times2)$ supercell of Mg$_2$Si. 

In the Mg$_2$Si$_{0.5}$Sn$_{0.5}$ model, Si atoms at (0, 0, 0), ($1/2$, 0, 0), (0, $1/2$, 0), (0, 0, $1/2$) are substituted for Sn atoms, in order to model a representative homogeneous alloy. The Mg$_2$Si$_{0.5}$Sn$_{0.5}$ supercell has a geometric strain of $+3.4\,\%$ compared to pure Mg$_2$Si, and $-3.2\,\%$ compared to Mg$_2$Sn. In the Mg$_2$Si$_{0.25}$Sn$_{0.75}$ model, six Si atoms are substituted for Sn, leaving only two Si atoms (at the origin and at the center) unchanged. The supercell has a strain of $+5.1\,\%$ and  $-1.6\,\%$, compared to pure Mg$_2$Si and Mg$_2$Sn, respectively. The Mg$_2$Si$_{0.75}$Sn$_{0.25}$ has only two Sn atoms, at the origin and at the center of the supercell. This supercell has a strain of $+1.7\,\%$ and  $-4.8\,\%$ when compared to pure Mg$_2$Si and Mg$_2$Sn, accordingly. All lattice parameters of the alloy models are scaled according to Vegard's law.

Thermoelectric materials operate at elevated temperatures, however, the calculated electronic structure conveys only the $T = 0$\,K ground state property. A recent study reported the explicit phonon-induced electronic structure calculations for Pb-based chalcogenides,\cite{skelton2014} and found that a high temperature (investigated up to about 600\,K) introduces variations in the eigenvalues of the order of 0.1\,eV. Noting that high operating temperatures can have a direct influence on the band degeneracies in these thermoelectric alloys, the explicit treatment of phonon-electron coupling has not been not taken into account in this work. Instead, we approach this problem by modelling the temperature effect via a statistical treatment of temperature-broadened electron occupancies in our band structure calculations at specified temperatures.

According to the Fermi-Dirac statistics, an elevated temperature results in a broadened distribution of electrons. To include the effect of temperature in our band structure calculations, we employ a Gaussian smearing function to approximate this broadening effect,\cite{devita1992} and use a Gaussian normal distribution to represent weight in momentum-energy space. We then use a convolution between the two distribution functions to obtain a temperature-broadened projected electronic band structure.

Specifically, once the ground state projected band structures are obtained, the additional calculations are performed in two steps. Firstly, the band occupancy is projected to a normal distribution along the energy axis (Eq.\,\ref{gaussian-smear1}). Here, the product of the Boltzmann constant and absolute temperature ($k_{\rm B}T$) is used as a broadening parameter ($\sigma$). Secondly, the calculated distribution is convoluted with a Gaussian smearing function (Eq.\,\ref{gaussian-smear2}). The final projection (Eq.\,\ref{gaussian-smear3}) is cast upon the energy axis at each {\bf k}-point, showing temperature-dependent broadening of the band occupancy. 

\begin{equation}
\label{gaussian-smear1}
\Phi = \frac{\exp{(-\frac{E^2}{2\sigma^2})}}{\sigma\sqrt{2\pi}} \quad,
\end{equation}
\begin{equation}
\label{gaussian-smear2}
f\left( \frac{E - \mu}{k_{\rm B}T}\right) = \frac{1}{2} \left(1-erf\left(\frac{E-\mu}{k_{\rm B}T}\right)\right)  \quad,
\end{equation}
\begin{equation}
\label{gaussian-smear3}
w(E, T) = \omega f\circ\Phi \quad.
\end{equation}

Here $\Phi$ denotes the projection of individual weight upon the energy ($E$) axis, $f\left( \frac{E - \mu}{k_{\rm B}T}\right)$ shows the Gaussian smearing function, where $\mu$ denotes the eigenvalue, and finally, a convoluted-weighted projection, $w$, is defined by the convolution of the Gaussian smearing and Gaussian distribution functions, multiplied by the calculated weight $\omega$. 

In our finite temperature approach, the broadening parameter $k_{\rm B}T$ is about  0.07\,eV at 800\,K in accordance with the variance range of the explicit phonon-induced band structure calculations.\cite{skelton2014}

\section{Results and discussion}
The calculated lattice parameters of Mg$_2$Si and Mg$_2$Sn are listed in Tab.\,\ref{table-lattice}. The PBE functional exhibits good agreement with experiment (within 2\,\%). The HSE06 shows a very similar result to the PBEsol for Mg$_2$Si, and performs slightly better for Mg$_2$Sn. The HSE06 functional is not frequently used for geometric optimization calculations, due to the significant computational cost which varies exponentially with increasing size of the system. The HSE06 functional describes the atomic geometry as well as the widely used efficient functionals in Tab.\,\ref{table-lattice}. 

\begin{table}[tb!]
\centering
\caption{Calculated primitive lattice parameter of Mg$_2$Si and Mg$_2$Sn (in \AA) obtained using various DFT $xc$ functionals. Their percentage differences in comparison to experimental results are shown in the parentheses.} 
\begin{ruledtabular}
\begin{tabular}{ccc}   
{\it xc} & Mg$_2$Si  &   Mg$_2$Sn \\\hline
PBE  		&  4.440 ($-$1.73)   &   4.742 ($-$0.80)    \\
PW91  		&  4.434 ($-$1.86)   &    4.738 ($-$0.88)   \\  
HSE06		&  4.419 ($-$2.20)   &    4.721 ($-$1.24)    \\
PBEsol  		&  4.419 ($-$2.20)   &   4.710 ($-$1.47)    \\
PBE+D2  		&  4.404 ($-$2.53)   &    4.720 ($-$1.26)   \\
PBE+D3  		&  4.397 ($-$2.68)   &    4.707 ($-$1.53)   \\ 
Expt.			&  4.518\footnote{Ref.\,\onlinecite{owen1923}}		&	4.780\footnote{Ref.\,\onlinecite{glazov1965}} \\
\end{tabular}
\end{ruledtabular}
\label{table-lattice}
\end{table}

A common approach to calculate the electronic structure of materials with a complex geometry is to first determine the ground state geometry with computationally cheaper (semi-)local functionals, and then to perform the more costly (albeit more accurate) band structure calculations at the determined geometry. This approach avoids significant computational cost and affords a more accurate description of certain material properties (e.g. more accurate band gaps). However, since the geometry found from the first step is not necessarily the equilibrium structure of the higher level functionals (though maybe very close), material properties that depend sensitively on the atomic structure may not be described as accurately as they could be. For instance, the lattice parameter of Mg$_2$Si calculated by the PBE is 4.440\,\AA, but HSE06 yields a 0.5\,\% smaller value. Thus, if we calculate the electronic structure from the two-step approach as described above, the results obtained by the HSE06 calculation will retain effectively a 0.5\,\% compressive strain. When geometric strain is considered an important factor to the property under investigation, as is in this study, it becomes pivotal to calculate both the lattice geometry and the electronic structure optimized under the same $xc$ functional. 

\begin{figure}[tb!]
\center
\includegraphics[width=0.70\textwidth]{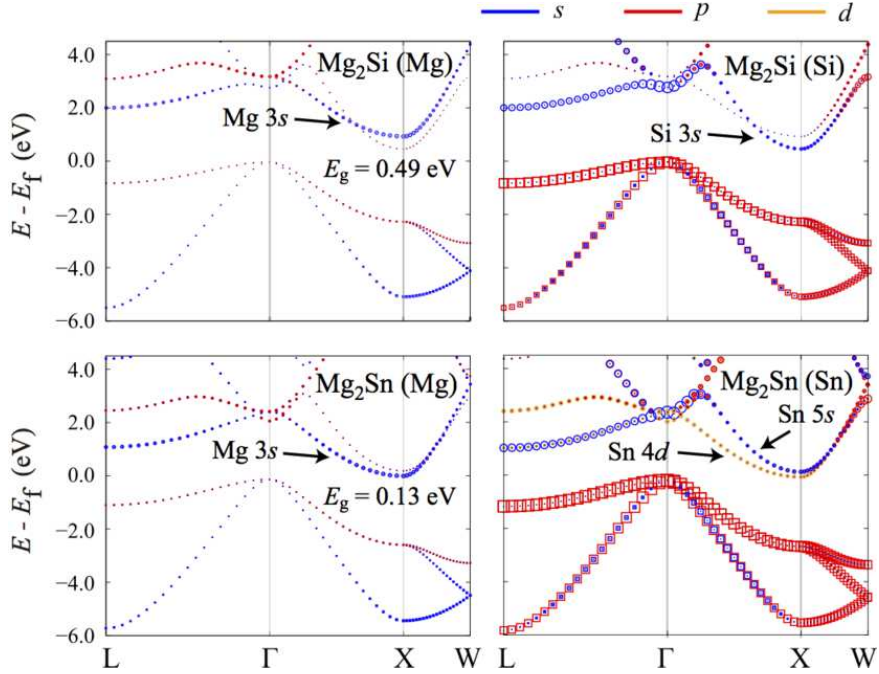} %0.45
\caption{(Color online) The projected band structure (atom species indicated in brackets) of Mg$_2$Si and Mg$_2$Sn as calculated using the HSE06 $xc$ functional. Circles are used for conduction bands, squares for valence bands. $E_{\rm g}$ is the indirect band gap between the $\Gamma$ and $X$ points. The size of the symbols indicate the magnitude of the weight of the orbital character at each {\bf k}-point. The results capture the Sn 4$d$ states in the low-lying heavy band. }
\label{figure-projband}
\end{figure}

The projected band structures of Mg$_2$Si and Mg$_2$Sn are shown in Fig.\,\ref{figure-projband}. Both compounds exhibit an indirect band gap between the conduction band minimum (CBM) at the $X$ point and the valence band maximum (VBM) at the $\Gamma$ point. The band gap of Mg$_2$Si is 0.49\,eV as calculated using the HSE06 functional. A $GW$ calculation gave 0.45\,eV,\cite{arnaud2000} and a spin-orbit coupled all-electron calculation with PBEsol gave 0.39\,eV.\cite{sharma2014} Though it is unseen from the graph, the Mg 3$s$ states has an inherent degeneracy of two since a primitive unit cell has two Mg atoms in symmetrically equivalent positions. 

The narrow band gap and conduction band degeneracy makes Mg$_2$Si alone an attractive thermoelectric material, yielding a $ZT$ value of 0.75 at 750\,K, when doped with 0.15\,\% Bi.\cite{bux2011} The compound Mg$_2$Sn has a narrower band gap of 0.13\,eV (as calculated using the HSE06 functional), which explains why Mg$_2$Sn displays relatively less thermoelectric power at high temperature compared to Mg$_2$Si. In both Mg$_2$Si and Mg$_2$Sn,  the CBM at the $X$ point has inherently high degeneracy, not only due to the crystalline symmetry, but also due to the closely located eigenvalues of Mg 3$s$, Si 3$s$, Sn 5$s$ and Sn 4$d$ states, which at high temperature may become effectively degenerate. For an optimal composition of Sn dopant, the degeneracy of conduction bands at the $X$ point in Mg$_2$Si$_{1-x}$Sn$_{x}$ solid solutions can be maximized, resulting in greater enhancement of the band valley degeneracy. 

The closely positioned conduction bands are distinguished in terms of their effective masses. A theoretical study of Mg$_2$Si$_{1-x}$Sn$_{x}$ showed that the heavy and light bands converge to form effectively degenerate states for $x$=0.625, with a calculated lattice parameter of 6.64\,\AA, which is roughly 4.5\,\% expanded relative to that of Mg$_2$Si, and 2.4\,\% compressed relative to that of Mg$_2$Sn, coinciding with the prediction by Vegard's law (6.63\,\AA).\cite{tan2012} 

\begin{figure}[tb!]
\center
\includegraphics[width=0.80\textwidth]{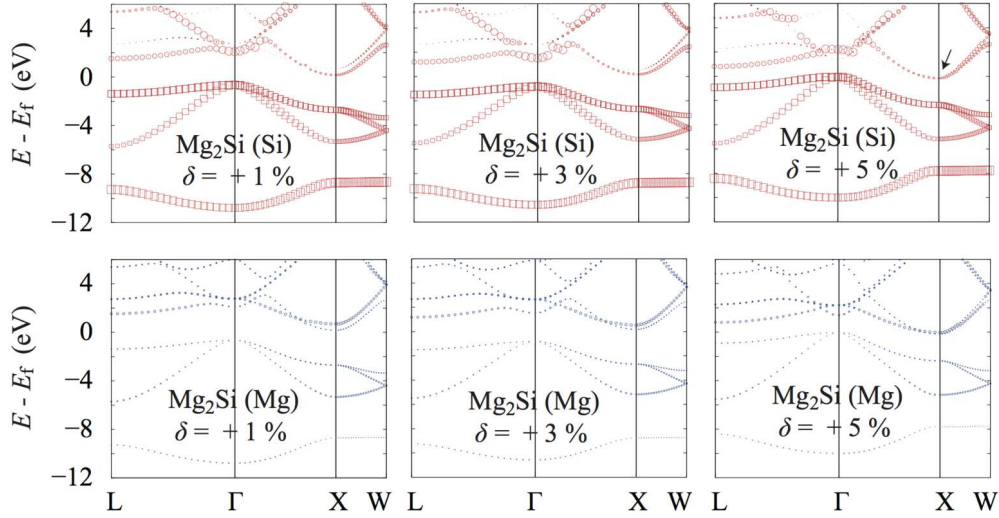} %0.48
\caption{(Color online) The projected band structures (the atom species indicated in brackets) of Mg$_2$Si, as calculated using the HSE06 $xc$ functional for various tensile strains ($\delta$). The weights of the orbital character are not separately shown here, only the sum is shown and indicated by the size of the symbols. The low-lying conduction bands at the $X$ point become degenerate when sufficiently large expansive lattice strain is present (indicated by an arrow).}
\label{figure-varylattice-mg2si}
\end{figure}

For Mg$_2$Si, the heavier conduction band consists mainly of Mg 3$s$ character, while the lighter band shows Si 3$s$ character. Interestingly, Mg$_2$Sn shows the presence of Sn 4$d$ states in the heavy band. The role of the delocalized $d$ states in enhancing the thermoelectric transport efficiency is still under active debate.\cite{ahmad2006,heremans2008,hsu2014} In the present study, the heavy $d$ band of Sn takes on an important role as well. The Sn 4$d$ states hybridize with Mg 3$s$ states in the conduction band, thereby further intensifying the band valley degeneracy in the CBM. 

\begin{figure}[tb!]
\center
\includegraphics[width=0.80\textwidth]{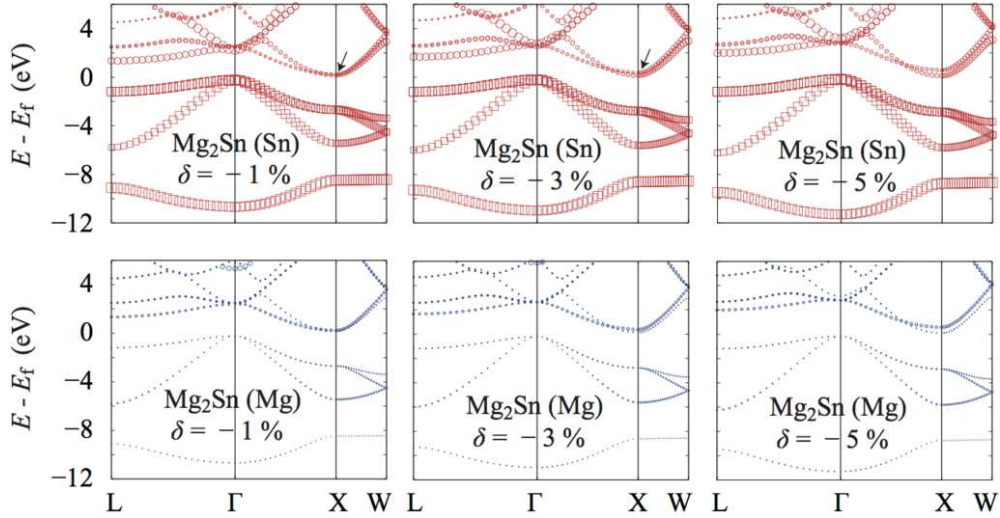} %0.48
\caption{(Color online) The projected band structures (the atom species indicated in brackets) of Mg$_2$Sn, as calculated using the HSE06 $xc$ functional for various values of compressive strain ($\delta$). The weights of the orbital character are not separately shown here, only the sum is shown and indicated by the size of the symbols. The low-lying conduction bands at the $X$ point become degenerate when an optimal range of compressive lattice strain is present (indicated by an arrow).}
\label{figure-varylattice-mg2sn}
\end{figure}

Turning to the effect of geometric lattice strain, we find that the varying lattice parameters significantly affects the position of the heavy and light bands (see Figs.\,\ref{figure-varylattice-mg2si} and \ref{figure-varylattice-mg2sn}). In particular, the projected band structures clearly show that geometrical strain may affect the light and heavy bands so that they overlap to form effectively degenerate levels. Mg$_2$Si shows an overlap for a 5\,\% expanded structure, and Mg$_2$Sn shows an overlap for a 1--3\,\% compressed structure. This behavior has been reported previously for both Mg$_2$Si and Mg$_2$Ge,\cite{krivosheeva2002} and here, we account for this behavior in Mg$_2$Sn for the first time. If we use Vegard's law to calculate the lattice parameter for $x=0.6$ in the Mg$_2$Si$_{1-x}$Sn$_{x}$ solid solution from experimental data of Liu {\it et al.},\cite{liu2012a,liu2013} the calculated lattice parameters also correspond to the optimal range of geometric lattice strain. These results therefore show that the increased degeneracy of the conduction bands is closely related to the degree of strain.

\begin{figure}[tb!]
\center
\includegraphics[width=0.49\textwidth]{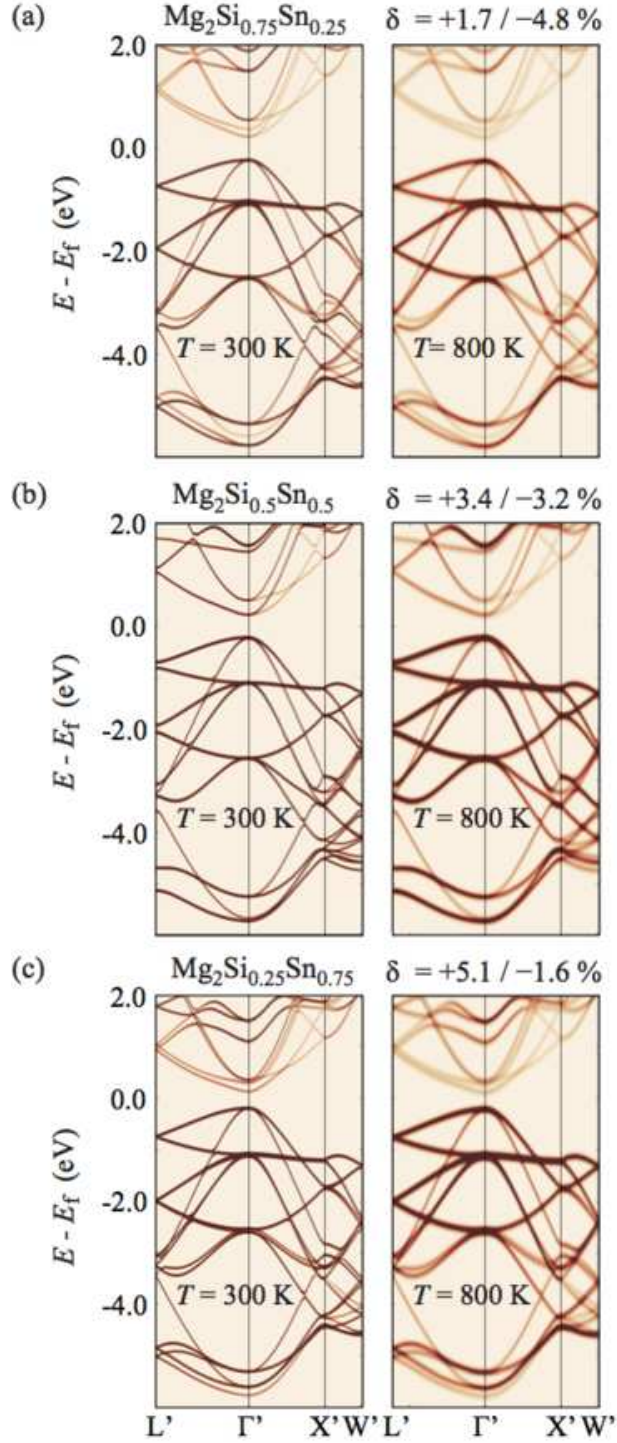} %0.35
\caption{(Color online) The calculated band structure of Mg$_2$Si$_{1-x}$Sn$_x$ solid solutions. The effect of finite temperature is calculated by convolution between a Gaussian smearing function and a Gaussian normal distribution function (see Sec.\,2 for details). The magnitude of the strain ($\delta$) in the sublattices is expressed as percentage difference with respect to pure Mg$_2$Si and Mg$_2$Sn, respectively. The lattice parameters of the supercells are calculated by Vegard's law. The notation of the reciprocal vectors are modified because the Brillouin zone of the supercell is different from that of the primitive cell.}
\label{figure-mss-band}
\end{figure}

We now address the band structures of our solid solution models of Mg$_2$Si$_{1-x}$Sn$_{x}$, as shown in Fig.\,\ref{figure-mss-band}. In a homogeneous solid solution of Mg$_2$Si$_{1-x}$Sn$_{x}$, both Si and Sn sub-lattices experience expansive and compressive strain, respectively. Therefore it is also important to investigate the electronic structure of these MSS solid solutions to determine the accurate position of degenerate levels with the HSE06 functional, having the varying degree of lattice strain in mind. 

Considering the temperature-broadened electronic band structure of Mg$_2$Si$_{0.25}$Sn$_{0.75}$ in Fig.\,\ref{figure-mss-band}, there exists multiple adjacent low-lying bands around the CBM. This can be explained in terms of the geometric strain. The lattice parameter of the solid solution model is about 5.1\,\% larger than pure Mg$_2$Si, and is about 1.6\,\% smaller than pure Mg$_2$Sn; where the low lying bands of the pure bulk models showed enhancement of the band valley degeneracy in previous results. 

The formation of the solid solution at this value of $x$ yields the optimal ranges of sublattice strain, where the band valley becomes maximally degenerate. Comparing the separation between the low-lying bands around the CBM to those for Mg$_2$Si$_{0.5}$Sn$_{0.5}$ and Mg$_2$Si$_{0.75}$Sn$_{0.25}$, it can be seen that for the latter structure, these bands are, relatively, rather far apart. This is because the lattice strain imposed by forming this solid solution deviates from the optimal ranges found for the pure compounds. On the other hand, when the geometric strain falls within the optimal range, the low-lying bands are close to each other and form effectively degenerate levels.

\begin{figure}[tb!]
\center
\includegraphics[width=0.75\textwidth]{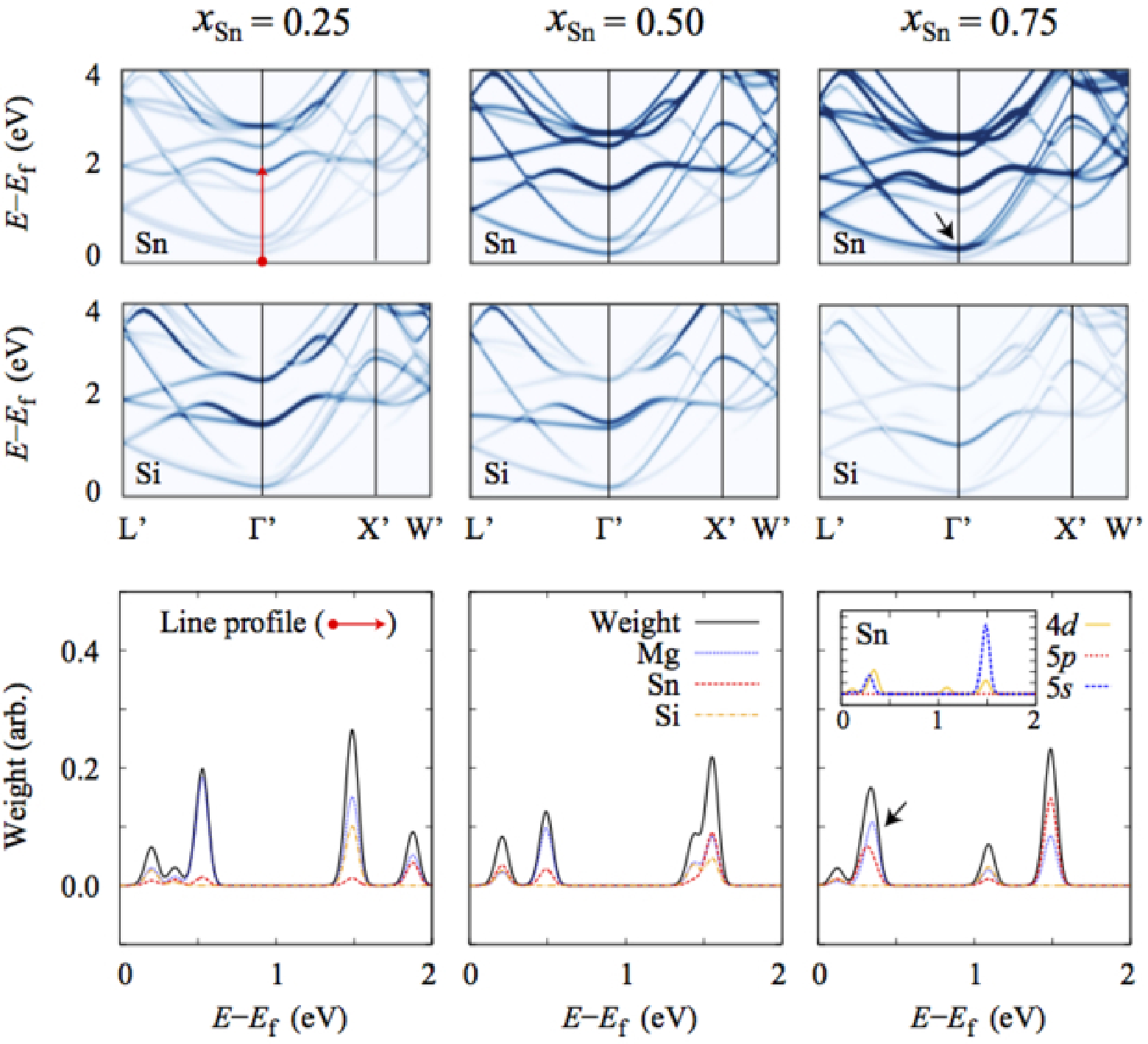} %0.48
\caption{(Color online) Upper plots: The finite-temperature induced projected band structure (atom species indicated) for various concentrations with finite temperature $T = 800$\,K. With $x = 0.75$ in Mg$_2$Si$_{1-x}$Sn$_x$ solid solutions, the degree of `effectively degenerate states' at the conduction band valley is greatest (indicated by the arrow). Lower plots: Line profiles along the energy axis (at the $\Gamma$ point). The orbital characters of Sn is shown in the inset.  }
\label{figure-mss-separate}
\end{figure}

Based on the calculated results, we attempt to explain the chemical and physical origins of the enhanced degeneracy in the conduction bands of Mg$_2$Si$_{1-x}$Sn$_{x}$ (MSS) alloys. We first discriminate the distinct contributions from Si and Sn from the projected electronic band structures (see Fig.\,\ref{figure-mss-separate}). In both MSS solid solution models, Sn persistently contributes to both heavy and light low-lying bands, while Si contributes less to the second low-lying bands. This leads us to believe that the chemical composition of the solid solution can be closely correlated with its enhanced band valley degeneracy. 

Besides this chemical origin, the conduction band states of the MSS alloys are also found to be significantly affected by the geometric strain of the sublattices. Compared to the constituent parent compounds Mg$_2$Si and Mg$_2$Sn, the Mg$_2$Si$_{0.25}$Sn$_{0.75}$ solid solution has an optimal range of the sublattice strain, where the conduction band states of pure constituents exhibit conduction band valley degeneracies, as well (as discussed in Figs.\,\ref{figure-varylattice-mg2si} and \ref{figure-varylattice-mg2sn}). With about $+5.1$\,\% expansive strain compared to pure Mg$_2$Si, and with $-1.6$\,\% compressive strain compared to pure Mg$_2$Sn, the $x_{\rm Sn} = 0.75$ alloy shows the most pronounced conduction band valley degeneracy. 

In addition to the geometric origin, we find the delocalized Sn 4$d$ and 5$s$ states significantly contribute to the enhanced band valley degeneracy (see inset of $x_{\rm Sn}=0.75$ results in Fig.\,\ref{figure-mss-separate}). We suggest x-ray absorption spectroscopy (XAS) can be a useful tool to investigate the supposedly increasing Sn 4$d$--Sn 5$s$ hybridization with respect to increasing proportion of Sn of Mg$_2$Si$_{1-x}$Sn$_{x}$ solid solution.

\section{Summary and conclusion}
The Mg-based alloy shows an excellent example where electronic band structure engineering lead to substantial strengthening of the performance. To conclude, we investigated the electronic structures of the alloy models ($x$=0.25, 0.50, and 0.75) as well as that of the constituent pure bulk materials (Mg$_2$Si and Mg$_2$Sn). We found that going beyond semi-local DFT (e.g. using the hybrid HSE06 functional) was necessary to accurately describe the band gap of both Mg$_2$Sn and its alloys. From the finite-temperature projected electronic band structures of these Mg-based alloys, we were able to elucidate and provide compelling evidence for the chemical and geometric origin of the observed conduction band valley degeneracy, and the significance of Sn 4$d$ states in the conduction band of the Mg$_2$Si$_{1-x}$Sn$_{x}$ alloy system. Finally, our finding contributes to an ongoing discussion about the electronic structure of MSS, using hybrid DFT temperature-broadened, orbital-projected band structure calculations. The enhanced thermoelectric character of the MSS solid solutions suggests a novel means to further improve the thermoelectric performance through electronic band structure engineering, and could well play a paramount role in the future discovery and design of novel and efficient thermoelectric materials for real devices.

\begin{acknowledgments}
We gratefully acknowledge support from the Basic Science Research Program by the NRF, (Grant No. 2014R1A1A1003415) and the Australian Research Council (ARC). This work was also supported by the third Stage of Brain Korea 21 Plus Project Division of Creative Materials. Computational resources have been provided by the Australian National Computational Infrastructure (NCI) and by the KISTI supercomputing center (KSC-2015-C3-009).
\end{acknowledgments}

\end{document}